\documentclass[12pt,showpacs,showkeys,amsmath,aps,prc,superscriptaddress]{revtex4}

\usepackage{amsfonts} 
\usepackage{mathrsfs}
\usepackage{subfigure}
\usepackage{supertabular}
\usepackage{color,graphicx}
\usepackage{multirow}
\usepackage[bookmarksopen]{hyperref}

\def  \p    {\pi}

\def  \m    {\mu}

\def  \th   {\theta}

\def  \ra   {\rightarrow}

\def  \del  {\partial}

\def  \bef  {\begin{figure}}
\def  \eef  {\end{figure}}
\def  \be   {\begin{equation}}
\def  \ee   {\end{equation}}
\def  \ba   {\begin{array}}
\def  \ea   {\end{array}}
\def  \bea  {\begin{eqnarray}}
\def  \eea  {\end{eqnarray}}
\def  \beq  {\begin{eqnarray}}
\def  \eeq  {\end{eqnarray}}
\def  \nn   {\nonumber}
\def  \bd   {\begin{displaymath}}
\def  \ed   {\end{displaymath}}
\def  \bse  {\begin{subequations}}
\def  \ese  {\end{subequations}}
\def  \bwt  {\begin{widetext}}
\def  \ewt  {\end{widetext}}

\def  \ba   {{\bf{a_1}}}

\begin{document}
\title{Non-Fermi liquid corrections to the neutrino mean free path in
dense quark matter}
\author {Kausik Pal}
\affiliation {High Energy Physics Division, Saha Institute of Nuclear Physics,
 1/AF Bidhannagar, Kolkata 700064, India.}
\affiliation{Department of Physics, Serampore College, Serampore 712201, India.}
\author {Abhee K. Dutt-Mazumder}
\affiliation {High Energy Physics Division, Saha Institute of Nuclear Physics,
 1/AF Bidhannagar, Kolkata 700064, India.}

\medskip

\begin{abstract}
We calculate the neutrino mean free path with non-Fermi liquid (NFL) corrections 
in quark matter from scattering and absorption processes for both degenerate 
and nondegenerate neutrinos.
We show that the mean free path decreases due to the non-Fermi liquid corrections 
leading to $l_{mean}^{-1}\sim[......+ .... C_F^2\alpha_s^2\ln(m_D/T)^2]$.
These reduction results in higher rate 
of scattering leading to faster cooling of stars for nondegenerate neutrinos.
For degenerate case the corresponding mean free path is found to be 
much smaller than the core radius of the neutron star which
maintains the local thermodynamic equilibrium.

\end{abstract}
\vspace{0.08 cm}



\maketitle

\section{Introduction}

Recently, there has been a substantial effort to study the properties of 
cold and warm quark matter. Such studies are important to understand the
properties of the astrophysical compact objects like neutron stars
and pulsars. There has been lot of experimental efforts like Einstein
laboratory, ROSAT, CHANDRA and XMM, where various measurements are
performed to understand the properties of neutron stars \cite{boya01,boyanov01}. 

There is a possibility that at the core of neutron stars
the density may go upto $5\sim 6$ times the normal nuclear matter density where
the matter is not expected to be in the hadronic phase. In fact, under
such a scenario one expects that it would be more appropriate to describe
the core of such dense stars as degenerate quark matter \cite{boya01} 
which is our main interest in the present work.

It is known that the newly born neutron stars cool via the emissions of 
neutrinos and antineutrinos within few minutes involving neutron beta decay 
and its inverse reaction \cite{iwamoto80,iwamoto82,schafer04}. This cooling behavior is related to the transport properties of the medium or
thermodynamical quantities like specific heat, entropy etc. 
All these quantities are in turn related to the neutrino scattering rate
or mean free path (MFP). 

Such emissions can take place either by the direct 
\cite{iwamoto80,iwamoto82,schafer04} or by the modified URCA processes
\cite{burrow79,iwamoto82}. The direct URCA reactions can proceed 
in neutron rich matter if the ratio of the proton number density
to the total baryon number density exceeds a critical value which follows 
from the energy and momentum conservation properties \cite{lattim91}.
In modified URCA process another nucleon catalyzes the reaction
to occur under situations where the direct URCA reaction is forbidden.

For quark matter, similar emission processes proceed through the decay of 
d quark or the scattering of u-d quarks. Analogously these reactions are named 
as `quark direct URCA' processes
which have been studied in detail by Iwamoto \cite{iwamoto80,iwamoto82}. Our main
focus here is to calculate the neutrino MFP, as mentioned earlier, for cold and warm QCD matter. Previously MFP of quark matter 
was derived in Ref.\cite{iwamoto80} where the calculations were restricted to
the leading order by assuming free Fermi gas interactions. Here we plan to go
beyond leading order to include `plasma' or `quasi particle' effects arising
out of the interacting ground state. This, as we show, gives rise to 
logarithmic corrections to the MFP in contrast to what one expects
from the usual Fermi liquid theory (FLT) \cite{baym76,pal09}. 
Similar non-Fermi liquid behavior of various thermodynamical quantities 
have recently drawn significant attention.

For example, in \cite{boya01,ipp04}, the authors
have computed the leading contribution to the interaction part of specific heat 
and entropy when temperature is much smaller than the chemical potential
of quark matter. It has been shown, in case of specific heat, 
instead of ${\cal O}(T^3)$ term, an 
anomalous $T\ln T$ term appear as corrections to the leading order
which becomes dominant for temperature smaller than the Debye mass. 
This type of nonanalytic
terms are known as non-Fermi liquid (NFL) corrections
\cite{boya01,boyanov01,Tschafer04,schafer04,ipp04,tatsumi09}. 
Ref.\cite{Tschafer04} examined 
NFL effects in the normal phase of high density QCD matter both using the 
Dyson-Schwinger equation and the renormalization group theory. 
In \cite{tatsumi09} magnetic susceptibility has also been shown to receive 
correction ${\cal O}(T^2\ln T)$. 
Being motivated by these series of works we undertake the present 
investigation to estimate and see the consequences
of such effects in the case of neutrino MFP.

In FLT, quarks are treated as quasiparticles and their 
energy $(E)$ is regarded as a functional of the distribution function
\cite{baym76,pal09,pal_gse,pal_sps,pal_mgs,pal10}. FLT is restricted to the 
low-lying excitations near the Fermi surface, where the lifetime of
quasiparticles are long enough. Therefore, it is an important tool to study
the properties of nuclear (or quark) matter. 
In Ref.\cite{brown00}, it has been argued,
that the exchange of dynamically screened transverse gluons introduces 
infrared divergences in the quark self-energies that lead to the breakdown 
of the Fermi liquid description of cold and dense QCD in perturbation theory.
A detailed study of non-Fermi liquid aspects of the normal state was presented
in Ref.\cite{boyanov01}. There the spectral density, dispersion relation 
and width of quasiparticles with momenta near the Fermi surface were derived 
at $T=0$ by implementing a renormalization group resummation of the leading 
logarithmic infrared divergences associated with the emission of soft
dynamically screened transverse gluons \cite{boya01,boyanov01}.

We have already mentioned that such anomalous corrections
are ultimately connected to the absence of magnetic screening of gluons
{\em via.} Landau damping. One such calculations which we find to the most 
relevant for the present purpose was performed in \cite{schafer04}. It was 
shown that the emissivity receives logarithmic corrections is enhanced due to 
non-Fermi liquid effects. It might not be out of context here to recall 
two important works on neutrino mean free path in QED plasma. One is due 
to Tubbs and Schramm \cite{tubb75} and the other is done by 
Lamb and Pethick \cite{lamb76}. In \cite{tubb75}, the resultant mean free
path was calculated in the neutranized core and 
just outside the core. It is concluded in \cite{lamb76}, 
that neutrino degeneracy reduces the neutrino mean free path which suggest
that neutrino may flow out of the core rather slowly.

In our work we show that the neutrino mean free 
path receive logarithmic corrections 
where the dressed gluon propagator is used instead of bare propagator. 
Infact, we generalize the \cite{lamb76,tubb75,iwamoto82} results by
incorporating the NFL corrections for the quark matter. The corrections
to the MFP for degenerate and non-degenerate neutrinos, as we shall see,
will involve different powers of $\alpha_s$.

It is well known that in the interior of the star the neutrino 
absorption and neutrino scattering can take place. These two types of mean 
free paths are related to different physical phenomenon. Absorption mean free path 
gives the transport properties, which characterizes the rate of momentum transport.
This also enters into the expression for the diffusion coefficient. While, the
scattering mean free paths are related to the relaxation time, which 
characterizes the rate of change of the neutrino distribution function 
\cite {iwamoto82}. To obtain the total mean free path 
one can write \cite{sagert06},
\beq
\frac{1}{l_{mean}^{total}}&=&\frac{1}{l_{mean}^{abs}}+\frac{1}{l_{mean}^{scatt}}
\eeq

\section{Mean free path}

In this section we calculate neutrino mean free paths in quark matter
including NFL corrections both for degenerate and nondegenerate neutrinos.
When the neutrino
chemical potential ($\mu_{\nu}$) is considered to be much larger than the temperature $(T)$, the neutrinos become degenerate and for 
nondegenerate neutrinos $\mu_{\nu}\ll T$. In our model the Lagrangian density 
is described by \cite{iwamoto80} 
\beq\label{lagrang}
{\cal L}_{Wx}(x)= \frac{G}{\sqrt 2}l_{\mu}(x){\cal J}^{\mu}_W(x)+H.C.
\eeq
where the weak coupling constant is $G \simeq 1.166\times 10^{-11}$ in MeV 
units, and $l_{\mu}$ and ${\cal J}^{\mu}_W$ are the lepton and hadron 
charged weak currents, respectively. The weak currents are
\beq
l_{\mu}(x)&=&{\bar e}\gamma_{\mu}(1-\gamma_5)\nu_e
+{\bar \mu}\gamma_{\mu}(1-\gamma_5)\nu_\mu+.... ,\\
{\cal J}^{\mu}_W(x)&=&\cos\th_c {\bar u}\gamma^{\mu}(1-\gamma_5)d
+\sin\th_c {\bar u}\gamma^{\mu}(1-\gamma_5)s+.... ;
\eeq
where $\th_c$ is the Cabibbo angle $(\cos^2 \th_c\simeq 0.948)$ \cite{barash80}.

The mean free path is determined by the quark 
neutrino interaction in dense quark matter {\em via.} weak processes. 
We consider the simplest $\beta$ decay reactions; the absorption 
process
\beq\label{dir}
d+\nu_{e}\rightarrow u+e^-
\eeq
and other is its inverse relation
\beq\label{inv}
u+e^-\rightarrow d+\nu_{e}
\eeq

The neutrino mean free path is related to the total interaction rate due 
to neutrino emission averaged over the initial quark spins and summed over 
the final state phase space and spins. It is given by \cite{iwamoto82}

\beq\label{mfp01}
\frac{1}{l_{mean}^{abs}(E_{\nu},T)}&=&\frac{g}{2E_{\nu}}\int\frac{d^3p_d}{(2\p)^3}
\frac{1}{2E_d}\int\frac{d^3p_u}{(2\p)^3}\frac{1}{2E_u}\int\frac{d^3p_e}{(2\p)^3}
\frac{1}{2E_e}(2\pi)^4\delta^4(P_d +P_{\nu}-P_u -P_e)\nn\\
&&|M|^2\{n(p_d)[1-n(p_u)][1-n(p_e)]-n(p_u)n(p_e)[1-n(p_d)]\},
\eeq
where, $g$ is the spin and color degeneracy, which in the present case is 
considered to be $6$. Here, $E$, $p$ and $n_p$ 
are the energy, momentum and distribution function for the corresponding particle. 
$|M|^2$ is the squared invariant amplitude averaged over 
initial $d$ quark spin and summed over final spins of $u$ quark and electron
as given by\cite{iwamoto82} 
\beq\label{mat01}
|M|^2 &=& \frac{1}{2}\sum_{\sigma_u,\sigma_d,\sigma_e}|M_{fi}|^2
=64G^2\cos^2\th_c(P_d\cdot P_\nu)(P_u\cdot P_e)
\eeq
Here, we work with the two flavor system as the interaction involving strange 
quark is Cabibbo suppressed \cite{boya01,schafer04}.

\subsection {Degenerate Neutrinos}
We now consider the case of degenerate neutrinos 
{\em i.e.} when $\mu_{\nu}\gg T$ or in other words we consider trapped
neutrino matter. So in this case both the direct (Eq.\ref{dir}) and inverse 
(Eq.\ref{inv}) processes can occur and both the terms in Eq.(\ref{mfp01}) under curly brackets \cite{iwamoto82} are retained. Consequently, 
the $\beta$ equilibrium condition becomes $\mu_d+\mu_{\nu}=\mu_u+\mu_e$. 
Neglecting the quark-quark interactions and by using Eqs.(\ref{mfp01}) 
and (\ref{mat01}), for the mean free path one obtains
\beq\label{mfp_nd1}
\frac{1}{l_{mean}^{abs,D}}&=&\frac{3}{4\pi^5}G^2\cos^2\th_c \int d^3p_d\int d^3p_u
\int d^3p_e (1-\cos\th_{d\nu})(1-\cos\th_{ue})\nn\\
&\times&\delta^4(P_d +P_{\nu}-P_u -P_e)[1+e^{-\beta(E_{\nu}-\mu_{\nu})}]
n(p_d)[1-n(p_u)][1-n(p_e)]
\eeq
In the square bracket, the second term $e^{-\beta(E_{\nu}-\mu_{\nu})}$ is
due to the inverse process (Eq.\ref{inv}). Since the masses of $u$, $d$ quark and electron are very small, one can neglect the mass effect on the mean free path.
To carry out the momentum integration
we define $p\equiv |p_d+p_\nu|=|p_u+p_e|$ as a variable. Following the 
procedure, described by Iwamoto \cite{iwamoto82} one has
\beq\label{sinth}
\sin\th_{d\nu} d\th_{d\nu}&=&\frac{p dp}{p_f(d) p_f(\nu)}\\
\label{costh}
(1-\cos\th_{d\nu})(1-\cos\th_{ue})&\simeq&
\frac{p^4-2p^2 p_f^2 +p_f^4}{4p_f(d)p_f(\nu)p_f(u)p_f(e)}
\eeq
and 
\beq\label{dpd}
d^{3}p_d&=&2\pi \sin\th_{d\nu} d\th_{d\nu}p_d^2 dp_d\nn\\
&=&2\pi\frac{p_f(d)}{p_f(\nu)}p dp \frac{dp_d}{dE_d}dE_d\nn\\
&=&2\pi\frac{p_f(d)}{p_f(\nu)}p dp \frac{dp_d}{d\omega}d\omega\\
\label{dpu}
d^{3}p_u&=&2\pi\frac{p_f(u)p_f(e)}{p}dE_e\frac{dp_u}{d\omega}d\omega
\eeq
where we denote the single particle energy $E_{d(u)}$ as $\omega$. 
For the free case $dp/d{\omega}$ is the inverse quark velocity.
It is well known that this slope of the dispersion relation changes in matter
due to scattering from the Fermi surface and excitation of the Dirac vacuum.
The modified dispersion relation can be obtained by computing the on-shell 
one-loop self-energy. For quasiparticles with momenta close to the Fermi 
momentum, the one-loop self-energy is dominated by the soft gluon 
exchanges \cite{manuel00}. The quasiparticle energy $\omega$ 
satisfies the relation \cite{manuel00,manuel96}
\beq\label{qpeng}
\omega&=&E_p(\omega)+{\rm Re}\Sigma(\omega,p(\omega))
\eeq
where we have approximated only to the real part of self energy, since the 
imaginary part of $\Sigma$ turns out to be negligible compared to its 
real part \cite{ipp04,tatsumi09}. The detailed analysis can be obtained in
\cite{pisarski90,manuel96}. 

In the relativistic case, when the Fermi velocity $v_f$ is close to the 
velocity of light $c$, exchange of magnetic gluons become important. In the 
non-relativistic case it is suppressed by a factor $(v/c)^2$ with respect 
to exchange of electric gauge bosons and is usually neglected. The magnetic 
interaction, as stated before, is screened only dynamically and the 
problem remains for the 
static gluons \cite{manuel00,sarkar10}. Therefore to obtain a finite result,
a suitable resummation has to be performed \cite{blaziot96,blaziot97}.
The analytical expressions for one-loop quark self-energy can be written as
\cite{Tschafer04,manuel00,brown00,ipp04,tatsumi09}
\beq
\Sigma&=&\frac{g^2 C_F}{12\pi^2}(\omega-\mu)\ln\Big(\frac{m_D}{\omega-\mu}\Big)
+ i\frac{g^2 C_F}{12\pi}|\omega-\mu|,
\eeq
It exhibits a logarithmic singularity close to the Fermi surface
{\em i.e.} when $\omega \ra \mu$. Thus the long ranged character of the 
magnetic interactions spoils the normal Fermi-liquid behavior 
\cite{baym76,pal09,pal_gse,pal_sps,pal_mgs,pal10}. The breakdown of 
the Fermi liquid picture is associated with the vanishing of the 
discontinuity of the distribution function at the Fermi surface 
\cite{boyanov01,brown00}. This non-perturbative
nature of the self energy gives rise to the non-Fermi liquid behavior. 
Here, $m_D$ is a cut-off factor and should 
be an order of the Debye mass. Differentiating Eq.(\ref{qpeng}) with respect to
$p$, we obtain $\frac{dp(\omega)}{d\omega}$ at leading order in $\frac{T}{\mu}$
is 
\beq\label{dpdo}
\frac{dp(\omega)}{d\omega}&\simeq& \Big(1-\frac{\del}{\del\omega}
{\rm Re}\Sigma(\omega)\Big)\frac{E_p(\omega)}{p(\omega)}\nn\\
&=&\Big[1+\frac{C_F\alpha_s}{3\pi}\ln\Big(\frac{m_D}{T}\Big)\Big]
\frac{E_p(\omega)}{p(\omega)}
\eeq
where $\alpha_s$ is the strong coupling constant, 
$C_F=(N_c^2-1)/(2N_c)$ and $N_c$ is the color factor.
Using Eqs.(\ref{dpdo}), (\ref{mfp_nd1}) and (\ref{costh}-\ref{dpu}), 
the neutrino mean free path can be determined for two conditions. 
For $|p_f(u)-p_f(e)|\ge |p_f(d)-p_f(\nu)|$
\beq\label{mfp_cond1}
\frac{1}{l_{mean}^{abs,D}}&=&\frac{4}{\pi^3}G^2\cos^2\th_c
\frac{\mu_u^2 \mu_e^3}{\mu_\nu^2}
\Big[1+\frac{1}{2}\Big(\frac{\mu_e}{\mu_u}\Big)
+\frac{1}{10}\Big(\frac{\mu_e}{\mu_u}\Big)^2\Big]\nn\\
&\times&[(E_\nu-\mu_\nu)^2+\pi^2 T^2]
\Big[1+\frac{C_F\alpha_s}{3\pi}\ln\Big(\frac{m_D}{T}\Big)\Big]^2
\eeq

To derive above Eq.(\ref{mfp_cond1}), we use the result of integral
\cite{iwamoto82,shapiro_book}
\beq
\int_{0}^{\infty}{dE_d}\int_{0}^{\infty}{dE_u}\int_{0}^{\infty}{dE_e}
[1+e^{-\beta(E_{\nu}-\mu_{\nu})}] n(p_d)[1-n(p_u)][1-n(p_e)]
\delta(E_d+E_{\nu}-E_u-E_e)\nn\\
\simeq \frac{1}{2}[(E_\nu-\mu_\nu)^2+\pi^2 T^2]~~~~~~~~~~~~~~~~~~~~~~~~~~~~~~~~~~~~~~~~~~
\eeq

Similarly, for $|p_f(d)-p_f(\nu)|\ge |p_f(u)-p_f(e)|$, the corresponding 
expression for mean free path can be obtained by replacing 
$\mu_u\leftrightarrow \mu_d$ and $\mu_e\leftrightarrow \mu_\nu$ 
in Eq.(\ref{mfp_cond1}). Since quark and electron are assumed to be massless,
the chemical equilibrium condition gives $p_f(u)+p_f(e)=p_f(d)+p_f(\nu)$,
which we use to derive Eq.(\ref{mfp_cond1}).

The other major contribution to the mean free path arises from quark-neutrino
scattering. The neutrino scattering process from degenerate quarks are given by
\beq
q_{i}+\nu_e({\overline \nu_e})\ra q_{i}+\nu_e({\overline \nu_e})
\eeq
for each quark component of flavor $i (=u~{\rm or}~d)$.
The scattering mean free path of the neutrinos in degenerate case can 
be calculated similarly as evaluated by Lamb and Pethick in \cite{lamb76} for
electron-neutrino scattering. Assuming $m_{q_i}/p_{f_i}\ll 1$ and including 
the non-Fermi liquid correction through phase space, the mean free path 
is given by
\beq\label{mfp_scd}
\frac{1}{l_{mean}^{scatt,D}}&=&\frac{3}{16}n_{q_i}\sigma_{0}
\Big[\frac{(E_\nu-\mu_\nu)^2+\pi^2 T^2}{m_{q_i}^2}\Big]
\Big[1+\frac{C_F\alpha_s}{3\pi}\ln\Big(\frac{m_D}{T}\Big)\Big]^2\Lambda(x_i)
\eeq 
Here, $m_{q_i}$ is the quark mass. $C_{V_i}$ and $C_{A_i}$ is the vector and 
axial vector coupling constant was given in TABLE-II of Ref.\cite{iwamoto82}. 
In Eq.(\ref{mfp_scd})
if we drop the color factor and the second square bracketed term, we 
obtain results reported in \cite{lamb76} for dense and cold QED plasma with
$m_q$ replaced by $m_e$. In Eq.(\ref{mfp_scd}) the constants 
$\sigma_0 \equiv 4G^2 m_{q_i}^2/\pi$ \cite{tubb75} and $n_{q_i}$ is the 
number density of quark, is given by 
\beq
n_{q_i}&=&2\int\frac{d^3p}{(2\p)^3}\frac{1}{e^{\beta(E_{q_i}-\mu_{q_i})}}
\eeq
where $2$ is the quark spin degeneracy factor. Explicit form of 
$\Lambda(x_i)$ can be written as \cite{lamb76,iwamoto82} 
\beq
\Lambda(x_i)&=&\frac{4}{3}\frac{{\rm Min}(\mu_\nu,\mu_{q_i})}{\mu_{q_i}}
\Big[(C_{V_i}^2 + C_{A_i}^2)\Big(2+\frac{1}{5}x_i^2\Big)+2C_{V_i}C_{A_i}x_i\Big]
\eeq
and $x_i=\mu_{\nu}/\mu_{q_i}$ if $\mu_{\nu}< \mu_{q_i}$
and $x_i=\m_{q_i}/\mu_{\nu}$ if $\mu_{\nu}> \mu_{q_i}$.

\subsection {Nondegenerate Neutrinos}

We also derive mean free path for nondegenerate neutrinos 
{\em i.e.} when $\mu_{\nu}\ll T$. For nondegenerate 
neutrinos the inverse process (\ref{inv}) is dropped. Hence we neglect
the second term in the curly braces of Eq.(\ref{mfp01}). In this case, only 
those fermions whose momenta lie close to their respective Fermi surfaces
can take part in a reaction. It is to be mentioned here, if quarks are 
treated as free, as discussed in \cite{iwamoto82,duncan83,huang07}, 
the matrix element vanishes, 
since $u$, $d$ quarks and electrons are collinear in
momenta. The inclusion of strong interactions between quarks relaxes
these kinematic restrictions resulting in a nonvanishing squared matrix amplitude.
Since the neutrinos are produced thermally, we neglect the neutrino momentum
in energy-momentum conservation relation 
\cite{iwamoto82}. This is not the case for degenerate neutrinos where 
$p_{\nu}\gg T$ and therefore such approximation is not valid there.
By doing angular average over the direction of the 
outgoing neutrino, from Eq.(\ref{mat01}) the squared matrix element is given by 
\cite{schafer04}
\beq\label{mat02}
|M|^2 &=& 64G^2\cos^2\th_c p_f^2(V_d\cdot P_\nu)(V_u\cdot P_e)\nn\\
&=& 64G^2\cos^2\th_c p_f^2 E_{\nu}\mu_e \frac{C_F \alpha_s}{\pi}
\eeq
where $V=(1,v_f)$ is the four velocity. To calculate 
Eq.(\ref{mat02}) we have used the chemical equilibrium condition
$\mu_d=\mu_u+\mu_e$ and also the relations derived from Fermi liquid theory are
given by\cite{schafer04}
\beq
v_F=1-\frac{C_F\alpha_s}{2\pi}~;~~~~~~ \delta \mu = \frac{C_F\alpha_s}{\pi}\mu
\eeq

Putting $|M|^2$ in Eq.(\ref{mfp01}) we have,
\beq\label{mfp02}
\frac{1}{l_{mean}^{abs,ND}}&=&\frac{3C_F\alpha_s}{4\p^6}G^2\cos^2\th_c
\int{d^3p_d}\int{d^3p_u}\int{d^3p_e}\nn\\
&&\delta^4(P_d +P_{\nu}-P_u -P_e)n(p_d)[1-n(p_u)][1-n(p_e)]
\eeq
Neglecting the neutrino momentum in the neutrino momentum conserving 
$\delta$ function, the integrals can be decoupled into two parts. 
Following the procedure of Iwamoto \cite{iwamoto82}, the angular integral 
is given by
\beq\label{partA01}
\mathcal{A}&=&\int{d\Omega_d}\int{d\Omega_u}\int{d\Omega_e}\delta(p_d-p_u-p_e)
~=~\frac{8\pi^2}{\mu_d\mu_u\mu_e}
\eeq
and the other part
\beq\label{partB01}
\mathcal{B}&=&\int_{0}^{\infty}p_d^2\frac{dp_d}{dE_d}{dE_d}
\int_{0}^{\infty}p_u^2\frac{dp_u}{dE_u}{dE_u}
\int_{0}^{\infty}p_e^2{dE_e}\nn\\
&&\delta(E_d+E_{\nu}-E_u-E_e)n(p_d)[1-n(p_u)][1-n(p_e)]
\eeq

Changing the variables to $x_d=(E_d-\mu_d)\beta$, $x_u=-(E_u-\mu_u)\beta$ 
and $x_e=-(E_e-\mu_e)\beta$ and denoting the single particle energy $E_{u(d)}$ 
as $\omega$ we have from Eq.(\ref{partB01})
\beq\label{partB02}
\mathcal{B}=\int_{-\infty}^{\infty}{dx_d}{dx_u}{dx_e}
\frac{dp_d(\omega)}{d\omega}\frac{dp_u(\omega)}{d\omega}p_d^2 p_u^2 p_e^2
\delta(x_d+x_u+x_e+\beta E_{\nu})n(x_d)n(-x_u)n(-x_e)
\eeq
As the contribution dominates near the Fermi surfaces, extension of lower 
limit is a reasonable approximation \cite{shapiro_book,haensel01}.

Using Eq.(\ref{dpdo}) and perform the integration of Eq.(\ref{partB02}) following the procedure defined in \cite{tatsumi09,shapiro_book,haensel01}, we have
\beq\label{partB03}
\mathcal{B}&=&\mu_d^2\mu_u^2\mu_e^2~\frac{(E_{\nu}^2+\pi^2 T^2)}
{2(1+e^{-\beta E_{\nu}})}
\Big[1+\frac{C_F\alpha_s}{3\pi}\ln\Big(\frac{m_D}{T}\Big)\Big]^2
\eeq

Using Eqs.(\ref{mfp02}), (\ref{partA01}) and (\ref{partB03}), the mean free path
at leading order in $T/\mu$ is given by
\beq\label{mfp03}
\frac{1}{l_{mean}^{abs,ND}}&=&\frac{3C_F\alpha_s}{\pi^4}G^2\cos^2\th_c
~\mu_d~\mu_u~\mu_e~
\frac{(E_{\nu}^2+\pi^2 T^2)}{(1+e^{-\beta E_{\nu}})}
\Big[1+\frac{C_F\alpha_s}{3\pi}\ln\Big(\frac{m_D}{T}\Big)\Big]^2
\eeq
The first term is known from \cite{iwamoto82} and the additional terms
are higher order corrections to the previous results derived in the
present work.

For the scattering of nondegenerate neutrinos in quark matter, the expression of 
mean free path was given by Iwamoto\cite{iwamoto82}. We incorporate the anomalous
effect which enters through phase space modification
giving rise to
\beq\label{mfp_scnd}
\frac{1}{l_{mean}^{scatt,ND}}&=&\frac{C_{V_i}^2 + C_{A_i}^2}{20}n_{q_i}\sigma_{0}
\Big(\frac{E_{\nu}}{m_{q_i}}\Big)^2\Big(\frac{E_{\nu}}{\mu_i}\Big)
\Big[1+\frac{C_F\alpha_s}{3\pi}\ln\Big(\frac{m_D}{T}\Big)\Big]^2.
\eeq
Here, we have assumed $m_{q_i}/p_{f_i}\ll 1$ and the constants $\sigma_{0}$ and number density $n_{q_i}$ defined earlier.

\section{Summary and Conclusion}

In this work we have calculated, the neutrino mean free path with NFL corrections 
in quark matter for both degenerate and nondegenerate neutrinos. Contribution
to the total mean free path arises from absorption and scattering processes. 
We find, for absorption processes, corrections to the MFP is ${\cal O}(\alpha_s^2)$
for degenerate neutrinos and for non-degenerate case it is ${\cal O}(\alpha_s^3)$.
This is due to the fact, that
for non-degenerate neutrinos, quarks and electrons are collinear in 
momenta on their Fermi surfaces, for which phase space for direct URCA
processes vanishes. To get nonvanishing phase space which requires
that the Fermi momenta of quarks and electrons should satisfy triangular 
relation, one must take into account the interaction between the quarks.

Armed with the results of the previous sections, we now 
estimate the numerical values of the 
neutrino mean free paths in quark matter both for degenerate
and nondegenerate neutrinos. The mean free paths are listed in 
Table-{\ref{fp_comp}} in unit of meter where the neutrino energy 
$(E_{\nu})$ is set to be equal to the temperature $(T)$ and 
$m_q= 10$ MeV\cite{iwamoto82}.
Following ref.\cite{schafer04} we take the quark chemical potential 
$\mu_q \simeq 500$ MeV corresponding to densities $\rho_b\approx 6\rho_0$, 
$\rho_0$ is the nuclear matter saturation density. The electron chemical 
potential is determined by using the charge neutrality and beta equilibrium conditions which yields $\mu_e=11$ Mev. The other parameter is used here 
are same as \cite{schafer04}.

\begin{table}
\caption{Comparison of neutrino mean free path in quark matter for the case of
degenerate and nondegenerate neutrinos. Mean free path given in unit of meter,
without NFL correction and including NFL correction for the case of absorption
and scattering processes.}
\label{fp_comp}
\begin{center}
\begin{tabular}{|c|c|c|c|c|c|}\hline\hline
{} &
\multicolumn{2}{|c|}{Absorption} &
\multicolumn{2}{|c|}{Scattering}\\\hline
{} & Without Correction &  With Correction & Without Correction & With Correction\\\hline\hline
Degenerate     & 3.62    & 1.85  &   15.88   &   8.13 \\ \hline
Nondegenerate  & 7.68   & 3.93  &  1585.18  &  811.06\\ \hline\hline
\end{tabular}
\end{center}
\end{table}

From Table-{\ref{fp_comp}},
we see that the anomalous logarithmic terms dominate the
mean free path and reduces its value considerably. This reduced mean free 
path results in higher rate of scattering leading to higher 
rate of cooling of the compact stars for nondegenerate neutrinos. We find 
the relatively short neutrino mean free path for degenerate neutrinos. These
suggest that conditions are favorable for building up of the neutrino number in
the core and the neutrino distribution function may approach close to that
required for the enhancement of local thermodynamic equilibrium.


\end{document}